\documentclass[showpacs,preprintnumbers,amsmath,amssymb]{revtex4}

\usepackage{graphicx}
\usepackage{dcolumn}
\usepackage{bm}
\begin{document}
\author{M. Sajjad Athar}
\affiliation{Department of Physics, Aligarh Muslim University, Aligarh-202 002, India}
\email{sajathar@rediffmail.com}
\author{I Ruiz Sim\'o} 
\affiliation{Departamento de F\'{\i}sica Te\'orica and IFIC, \\
Centro Mixto Universidad de
Valencia-CSIC,\\
46100 Burjassot (Valencia), Spain}
\email{igruizsi@ific.uv.es}
\author{S. K. Singh}
\affiliation{Department of Physics, Aligarh Muslim University, Aligarh-202 002, India}

\author{M J Vicente Vacas}
\affiliation{Departamento de F\'{\i}sica Te\'orica and IFIC, \\
Centro Mixto Universidad de
Valencia-CSIC,\\
46100 Burjassot (Valencia), Spain}
\title{A study of nuclear effect in $F_3$ structure function in the deep inelastic $\nu(\bar\nu)$ reactions in nuclei}


\begin{abstract}
We study nuclear effect in the $F^A_3(x)$ structure function in the deep inelastic neutrino reactions on iron by taking into account Fermi motion, binding, target mass correction, shadowing and anti-shadowing corrections. Calculations have been done in a local density approximation using relativistic nuclear spectral functions which include nucleon correlations for nuclear matter. Results for $F^A_3(x)$ have been compared with the results reported at NuTeV and also with some of the older experiments reported in the literature.
\end{abstract}

\maketitle


\bigskip

\maketitle
With the recent development of accelerator-based neutrino experiments to look out for the oscillation searches has also necessitated the importance of measuring neutrino nucleus cross sections in the wide range of neutrino energies. As most of these experiments are in the few GeV energy region where the contribution to the cross section comes from the various processes the importance of precise knowledge of neutrino nucleus cross section has been realized. MINER$\nu$A using neutrinos from NUMI Lab., is going to perform cross section measurement in the neutrino energy region of 1-20GeV and with various nuclear targets like Carbon, Iron, Lead, etc. Recently NuSOnG experiment has been proposed to study the structure function in the deep inelastic region using neutrino sources. NuTeV collaboration~\cite{Tzanov} has reported the results on weak charged and neutral current induced (anti)neutrino processes on an iron target in the deep inelastic region. It has been emphasized that the  nuclear medium effect is important. In the case of deep inelastic scattering of (anti)neutrinos from nuclear targets, there are few calculations where the dynamical origin of the nuclear medium effect has been studied, and in some theoretical analysis, nuclear medium effect has been phenomenologically described in terms of a few parameters which are determined from fitting the experimental data of charged leptons and (anti)neutrino deep inelastic scattering from various nuclear targets. The differential scattering cross section for the deep inelastic scattering of (anti)neutrinos from unpolarized nucleons in the limit of lepton mass $m_l \rightarrow 0$, is described in terms of three structure functions, $F^\nu_1$($x$,$Q^2$), $F^\nu_2$($x$,$Q^2$) and $F^\nu_3$($x$,$Q^2$), where $x=\frac{Q^2}{2M\nu}=-\frac{q^2}{2M\nu}$ is the Bjorken variable, $\nu$ and $q$ being the energy and momentum transfer of leptons. In the asymptotic region of Bjorken scaling i.e. $Q^2 \rightarrow \infty$, $\nu \rightarrow \infty$, $x$ finite, all the structure functions depend only on the Bjorken variable $x$. In this scaling limit, $F^\nu_1(x)$ and $F^\nu_2(x)$ are related by the Callan-Gross relation leading to only two independent structure functions $F^\nu_2$($x$) and $F^\nu_3$($x$) which are determined from the experimental data on deep inelastic scattering of (anti)neutrinos in the asymptotic region. In this paper, we  study nuclear medium  effect on the nucleon structure function $F^A_3$($x$,$Q^2$) in iron. We use a theoretical spectral function~\cite{FernandezdeCordoba:1991wf} to describe the momentum distribution of nucleons in the nucleus. The spectral function has been calculated using the Lehmann's representation for the relativistic nucleon propagator and nuclear many body theory is used to calculate it for an interacting Fermi sea in nuclear matter. A local density approximation is then applied to translate these results to finite nuclei~\cite{Marco, Sajjad}. We have taken the effect of nuclear shadowing and anti-shadowing from the works of Kulagin and Petti~\cite{Petti}.  For the numerical calculations, NNLO parton distribution functions for the nucleons have been taken from the parametrization of Martin et al.(MSTW)~\cite{MSTW}. The NNLO evolution of the deep inelastic structure functions has been taken from the works of Moch et al.~\cite{Moch}. The target mass correction has been taken from the works of Schienbein et al.~\cite{Schienbein}. We have compared our results with nuclear medium effect with the experimental results reported by NuTeV~\cite{Tzanov} and CDHSW~\cite{Berge} Collaborations for a wide range of x and $Q^2$.
\begin{center}
\begin{figure}
\includegraphics[scale=0.6]{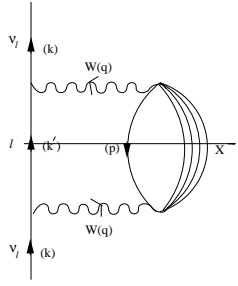}
\caption{Self-energy diagram of the neutrino in the nuclear medium
associated with the process of deep inelastic neutrino-nucleon
scattering. The imaginary part is calculated by cutting along the horizontal line and applying the Cutkosky rule for putting the particles on the mass shell.}
\end{figure}
\end{center}
 In the present formalism the neutrino nuclear cross sections are obtained in terms of neutrino self energy $\Sigma(k)$ in the nuclear medium which is defined as $ d \sigma= - \frac{2 m_\nu}{|\bf k|} \;
\hbox{Im} \; \Sigma \; d^3 r$, where $\Sigma(k)$ in nuclear matter corresponding to Fig.1 is given by,
\begin{eqnarray}\label{Sigma}
\Sigma (k) =&& (-i)\frac{G_F}{\sqrt{2}}\frac{4}{m_\nu}
\int \frac{d^4 k^\prime}{(2 \pi)^4} \frac{1}{{k^\prime}^2-m_l^2+i\epsilon} 
\left(\frac{m_W }{q^2-m_W^2}\right)^2 \; L_{\alpha \beta} ~~ \Pi^{\alpha \beta} (q)\,,
\end{eqnarray}
where $L_{\alpha\beta}$ is the leptonic tensor and $\Pi^{\alpha \beta} (q)$ is the $W$ self-energy in the nuclear medium and is written as:
\begin{eqnarray}
\Pi^{\alpha \beta} (q) &=& ( -i ) \; \int \frac{d^4 p}{(2 \pi)^4} \sum_{s_p, s_i} \prod^n_{i = 1}\int \frac{d^4 p'_i}{(2 \pi)^4}\prod_l i G_l (p'_l) \prod_j \; i D_j (p'_j) \left( \frac{-G_F m_W^2}{\sqrt{2}} \right)\nonumber\\
&&\times\langle X \mid J^{\alpha} \mid N \rangle \langle X \mid J^{\beta} \mid N \rangle{^*}(2 \pi)^4 \delta^4 (q + p - \Sigma^n_{i = 1} p'_i)
\end{eqnarray}
In the above expression  $G_l (p'_l)$ and $D_j (p'_j)$ are respectively the nucleon and meson relativistic propagators in the final state which are taken as the standard free relativistic propagators. The  nucleon propagator $G(p)$ inside the nuclear medium is calculated by making a perturbative expansion of $G(p)$ in terms of free nucleon propagator $G^{0}(p)$. This relativistic nucleon propagator after summing over the ladder approximation give~\cite{Marco}:
\begin{eqnarray}
G(p_{0},{\bf p})&=&\frac{M}{E({\bf p})}\sum_{r}\frac{u_{r}(p)\bar u_{r}(p)}{p^{0}-E({\bf p})-\bar u_{r}(p)\sum(p^{0},{\bf p})u_{r}(p)\frac{M}{{\bf E(p)}}}
\end{eqnarray}
This allows us to write the relativistic nucleon propagator in a nuclear medium in terms of the Spectral functions of holes and particles and the expressions have been taken from the works of Fernandez de Cordoba and Oset~\cite{FernandezdeCordoba:1991wf}. We use local density approximation (LDA) and for a symmetric nuclear matter of density $\rho({\bf r})$, 
\begin{eqnarray*} 
4\int\frac{d^{3}p}{(2\pi)^{3}}\int_{-\infty}^{\mu}S_{h}(\omega,p,k_{F}({\bf r})) d\omega= \rho({\bf r}) \; ~~and~~ \; 4 \int d^3 r \;  \int \frac{d^3 p}{(2 \pi)^3} 
\int^{\mu}_{- \infty} \; S_h (\omega, p, \rho(r)) 
\; d \omega = A
\end{eqnarray*}
where $\rho(r)$ is the baryon density for the nucleus which is normalized to A.

The expression of $F^A_3 (x, Q^2)$ due to nuclear medium effect like Fermi motion and binding in the present model is given by~\cite{Sajjad}:
\begin{eqnarray}\label{F3A}
F^A_3 (x, Q^2)&=& 4 \int d^3 r \int \frac{d^3 p}{(2 \pi)^3} \frac{M}{E (p)}
\int^{\mu}_{- \infty} d p^0 S_h (p^0, p, \rho(r))\left(\frac{p_0 \gamma-p_z}{(p_0-p_z\gamma) \gamma}\right)F^N_3(x_N, Q^2)
\end{eqnarray}
where $\gamma$ is $
\gamma=\frac{q_z}{q^0}= \left(1+\frac{4M^2x^2}{Q^2}\right)^{1/2}$.

The average structure function $F_{3}^{N}(x)$ on isoscalar nucleon target is defined as
\begin{eqnarray*}\nonumber
F_{3}^{N}(x)=\frac{1}{2}\left(F^{\nu N}_{3} + F^{\bar\nu N}_{3}\right)=[ u_v(x) + d_v(x) + s(x)- {\bar s}(x) + c(x) - {\bar c}(x) + b(x) - {\bar b}(x)]
\end{eqnarray*} 
where $u_v(x)=u(x) - {\bar u}(x)$ and $d_v(x)=d(x) - {\bar d}(x)$ are the valence quark parton distributions. The parton distribution functions for the nucleons have been taken from the NNLO parametrization of Martin et al.(MSTW)~\cite{MSTW}.

We take into account the QCD corrections to the structure function $F_3$, following the works of Moch et al.~\cite{Moch}. The charged-current coefficient function 
$C_3^{\,-}$ to the second order in the strong coupling $\alpha_s$, i.e., the coefficients $c_{3,-}^{\,(l\leq 2)}$ in
\begin{equation}\label{C3min} 
  F_3^{W^+ + W^-} = \; C_3^{-} \otimes q_{\rm val}=\left(\delta(1-x)+ \frac{\alpha_s}{4\pi} c_{3,-}^{(1)}+(\frac{\alpha_s}{4\pi})^2c_{3,-}^{(2)}\right) \otimes q_{val}
\end{equation}
Here $q_{val}$ represents the total (flavor summed) valence quark
distribution of the hadron, $q_{val} = \sum_{i=1}^{N_f} (q_i - \bar{q}_i)$ where $N_f$ is the number of effectively mass-less flavours. 
$\otimes$ denotes the Mellin convolution. We have taken five quark flavors (u to b) into account. The coefficients $c_{3,-}^{(1,2)}$ are taken from the Ref.~\cite{Moch} and references there in.
\begin{figure}
\includegraphics[scale=0.54]{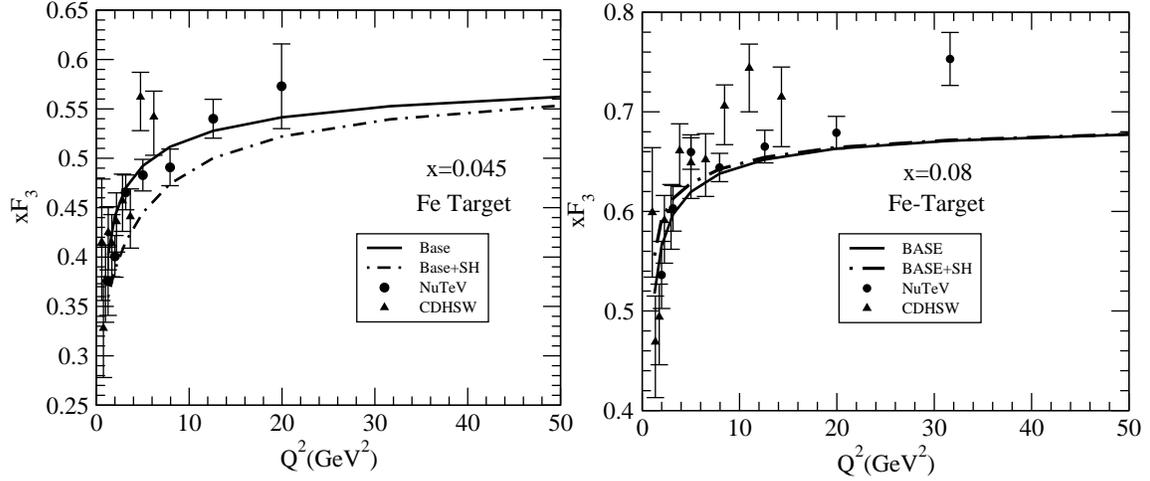}
\caption{Results for $F^A_{3} (x)$ vs $Q^2(GeV^2)$ in $^{56}Fe$ at x=0.045 and x=0.08. The results have been shown for $F^A_{3} (x)$ calculated using Eq.~(\ref{F3A})(Base) and with target mass correction and shadowing $\&$ anti-shadowing effect(dashed line). The experimental points are NuTeV~\cite{Tzanov} and CDHSW ~\cite{Berge} results.}
\end{figure}

\begin{figure}
\includegraphics[scale=0.54]{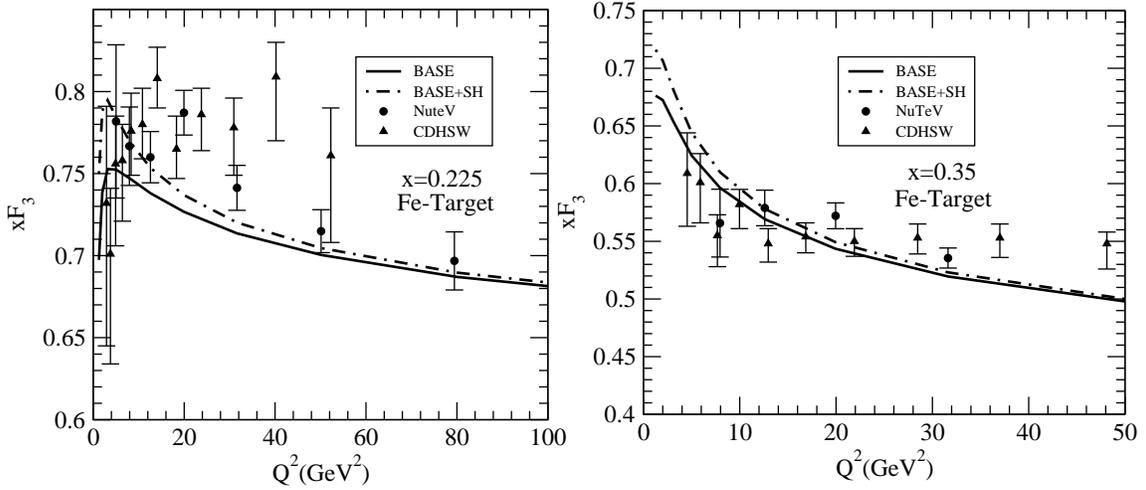}
\caption{Results for $F^A_{3} (x)$ vs $Q^2(GeV^2)$ in $^{56}Fe$ at x=0.225 and x=0.35.}
\end{figure}

\begin{figure}
\includegraphics[scale=0.54]{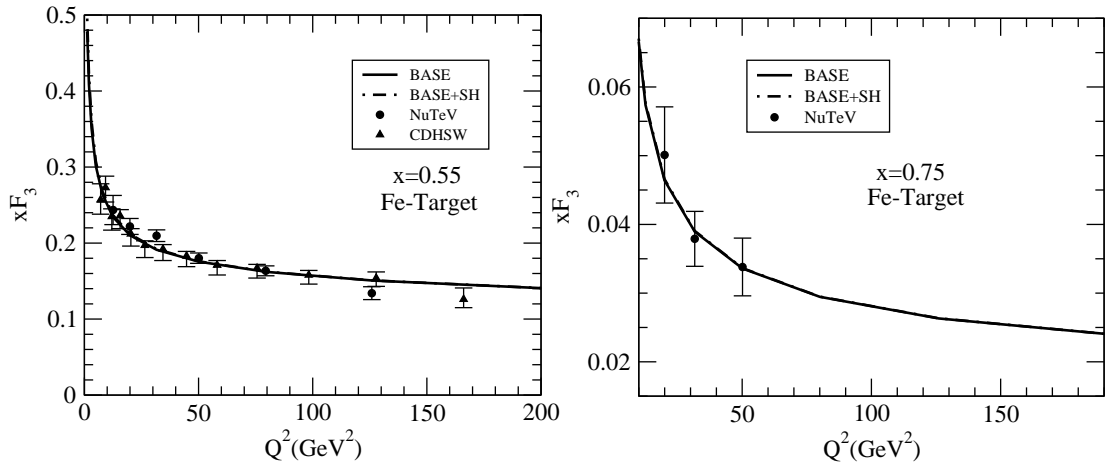}
\caption{Results for $F^A_{3} (x)$ vs $Q^2(GeV^2)$ in $^{56}Fe$ at x=0.55 and x=0.75. }
\end{figure}
The modification due to the target mass correction in $F_3$ structure function has been taken from the works of Schienbein et al.\cite{Schienbein}, which has been approximated by
\begin{eqnarray*}
F_{3}^{{\rm TMC}}(x,Q^{2}) \simeq \frac{x}{\xi r^{2}} F_{3}^{(0)}(\xi)
\bigg[1-\frac{\mu x \xi}{r} (1-\xi) \ln \xi \bigg]\, \; where \; r=\sqrt{1+\frac{4x^2M^2}{Q^2}}; \xi=\frac{2x}{1+r}; \mu=\frac{M^2}{Q^2}.
\end{eqnarray*}
Using Eq.(\ref{F3A}), we have studied the ratio R($x$,$Q^2$)=$\frac{F^A_3(x,Q^2)}{AF^N_3(x,Q^2)}$ with respect to $x$ at the different values of $Q^2$, and find that there is a suppression for $x \le x_{min}$=0.7, beyond which we get an enhancement. This value of ${x}_{min}$ decreases with increase in $Q^2$. For $x > x_{min}$, the ratio increases very fast and becomes larger than unity as $x\rightarrow$ 1. This is mainly due to the  Fermi motion of the nucleons. In the region of 0.3$<x<$1, this behavior is very similar to that seen in the EMC effect with charged lepton and (anti)neutrino scattering from the nuclear targets in the analysis of the structure function $F^{A,l}_2(x, Q^2)$ and $F^{A, \nu(\bar\nu)}_2(x,Q^2)$~\cite{Sajjad}.

Furthermore, we have taken into account the shadowing effect following the works of Kulagin and Petti~\cite{Petti} which is based on the Glauber-Gribov multiple scattering theory. At small x it is well known that the partonic distributions are dominated by sea quarks and gluons and an intermediate boson fluctuates into quark-antiquark pair which results in multiple scattering. It has been observed that the effect of shadowing decreases with increase in x and increases as one goes from low mass to heavy mass nuclei. It also decreases with increase in Q$^2$. We find that at x=0.45 and $Q^2=1-2GeV^2$, the increase in $F_3(x,Q^2)$ structure function is around 25-30$\%$ which reduces to $5-7\%$ at $Q^2=5-10GeV^2$, while at x=0.65 and $Q^2=1-2GeV^2$, the increase in $F_3(x,Q^2)$ structure function is around 70-75$\%$ which reduces to $25-30\%$ at $Q^2=5-10GeV^2$ and around 5$\%$ at $Q^2=20GeV^2$. 

The effect of shadowing-antishadowing has been studied by the prescription given by Kulagin and Petti~\cite{Petti} and we find that it is important at low x and low $Q^2$. For example at x=0.0001 to x=0.015, 
the reduction in $F_3(x,Q^2)$ structure function is around 40-50$\%$ at $Q^2$=1-2GeV$^2$ which becomes to $30-35\%$ at $Q^2=3-5GeV^2$, and around 10$\%$ for $Q^2$=20GeV$^2$. For x=0.05, the reduction in $F_3(x,Q^2)$ structure function is around 10$\%$ at $Q^2$=1-2GeV$^2$ which becomes to $8-10\%$ at $Q^2=3-5GeV^2$, and around 3-4$\%$ for $Q^2$=20GeV$^2$. While for x=0.08, we get an enhancement of around 8$\%$
at $Q^2$=1-2GeV$^2$ which becomes 2-3$\%$ for $Q^2$=3-5GeV$^2$. For x=0.1-0.3, the enhancement in $F_3(x,Q^2)$ structure function is around 10$\%$ at low $Q^2$ which dies out rapidly with the increase in $Q^2$.
 
In Figs.2-4, we have shown the results for $F^A_3(x,Q^2)$ with nuclear medium effect as a function of $Q^2$ for the various values of x and compared our results with the experimental results of NuTeV~\cite{Tzanov} and CDHSW~\cite{Berge} collaborations and found that the present results are in qualitative agreement with the experimental results. Therefore, to conclude, we find that nuclear medium effect plays an important role in the study of $F^N_3(x,Q^2)$ structure function in nuclei.

\end{document}